\documentclass[twocolumn,aps,prl,superscriptaddress,showpacs,amsmath,amssymb,floats]{revtex4}

\usepackage{amsmath}
\usepackage{amssymb}
\usepackage{graphicx}
\usepackage{CJK}
\usepackage{keyval}
\usepackage{dcolumn}
\usepackage{bm}
\usepackage{color}
\usepackage{txfonts}
\usepackage[]{sidecap}

\begin{document}

\begin{CJK*}{UTF8}{gbsn}

\title{Enhanced superconductivity and evidence for novel pairing in single-layer FeSe on SrTiO$_3$ thin film under large tensile strain}

\author{R. Peng}\author{X. P. Shen}\author{X. Xie}\author{H. C. Xu}\author{S. Y. Tan}\author{M. Xia} \author{T. Zhang}
\affiliation{State Key Laboratory of Surface Physics, and Department of Physics, Fudan University, Shanghai 200433, China}
\affiliation{Advanced Materials Laboratory, Fudan University, Shanghai 200433, People's Republic of China}

\author{H. Y. Cao}\author{X. G. Gong}

\affiliation{State Key Laboratory of Surface Physics, and Department of Physics, Fudan University, Shanghai 200433, China}
\affiliation{ Key Laboratory for Computational Physical Sciences (MOE), Fudan University, Shanghai 200433, China}

\author{J. P. Hu}
\affiliation{Beijing National Laboratory for Condensed Matter Physics,
Institute of Physics, Chinese Academy of Sciences, Beijing 100080, People's Republic of China}
\affiliation{Department of Physics, Purdue University, West Lafayette, Indiana 47907, USA}

\author{B. P. Xie}\email{bpxie@fudan.edu.cn}\author{D. L. Feng}\email{dlfeng@fudan.edu.cn}
\affiliation{State Key Laboratory of Surface Physics, and Department of Physics, Fudan University, Shanghai 200433, China}
\affiliation{Advanced Materials Laboratory, Fudan University, Shanghai 200433, People's Republic of China}

\date{\today}
\begin{abstract}

Single-layer FeSe films with extremely expanded in-plane lattice constant of 3.99$\pm$0.02$~\AA$ are fabricated by epitaxially growing FeSe/Nb:SrTiO$_3$/KTaO$_3$ heterostructures, and studied by \textit{in situ} angle-resolved photoemission spectroscopy. Two elliptical electron pockets at the Brillion zone corner are resolved with negligible hybridization between them, indicating the symmetry of the low energy electronic structure remains intact as a free-standing single-layer FeSe, although it is on a substrate.
The superconducting gap closes at a record high temperature of 70~K for the iron based superconductors. Intriguingly, the superconducting gap distribution is anisotropic but nodeless around the electron pockets, with minima at the crossings of the two pockets. Our results put strong constraints on the current theories, and support the coexistence of both even and odd parity spin-singlet pairing channels as classified by the lattice symmetry.

\end{abstract}


\maketitle

\end{CJK*}

\begin{figure}[t]
\includegraphics[width=8.6cm]{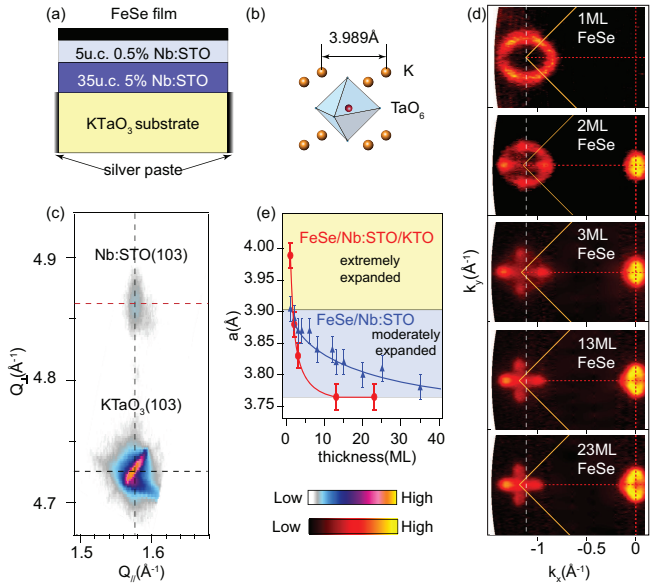}
\caption{(color online). (a) Schematic illustration of the heterostructure. (b) Cubic unit cell of KTaO$_3$ substrate. (c)	X-ray diffraction reciprocal space map of the grown heterostructure around the (103) Bragg reflections. (d) Thickness-dependent photoemission intensity maps at the Fermi energy (E$_F$) for FeSe/Nb:SrTiO$_3$/KTaO$_3$. The intensity was integrated over a window of [E$_F$-10 meV, E$_F$+10 meV]. (e) In-plane lattice of FeSe as a function of FeSe thickness deduced from the Brillion zone measured by ARPES. The data of FeSe/Nb:SrTiO$_3$ is reproduced from ref.~\onlinecite{FeSeTan}.}
\label{film}
\end{figure}

\begin{figure*}[t]
\includegraphics[width=17cm]{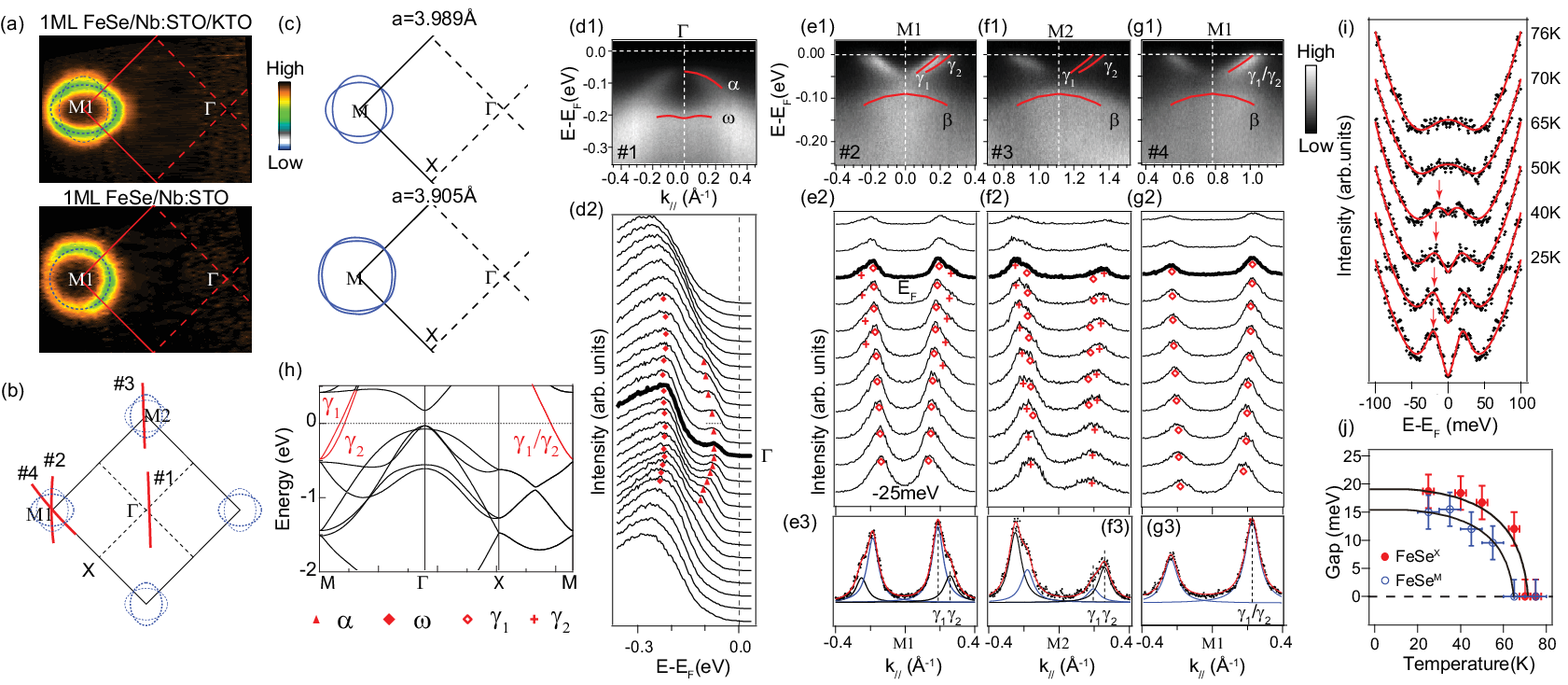}
\caption{(color online). (a) Photoemission intensity map of FeSe$^X$, compared with that of FeSe$^M$ which is reproduced from ref.~\onlinecite{FeSeTan}. The intensity was integrated over a window of [E$_F$-10 meV, E$_F$+10 meV]. The the measured Fermi surface sheets are shown by the dashed curves.
(b) Four-fold symmetrized sketch of the Fermi surface sheets observed in panel (a) for FeSe$^X$. Cuts $\#$1, $\#$2, $\#$3, $\#$4 are indicated in the Brillion zone.
(c) Calculated Fermi surfaces of single-layer FeSe based on DFT with the in-plane lattice constant of 3.989$\AA$ and 3.905$\AA$, to model FeSe$^X$ and FeSe$^M$, respectively. The extra 0.12 e$^-$ per Fe are included by shifting the chemical potential \cite{GXG}.
(d1) Photoemission intensity along cut $\#$1 through $\Gamma$ as indicated in panel (b), and the corresponding (d2) energy distribution curves (EDCs). (e1) Photoemission intensity along cut $\#$2, (e2) the corresponding MDCs, and (e3) MDC at E$_F$ fitted by Lorentzian peaks.
(f1)-(f3) are the same as (e1)-(e3), but along cut $\#$3.
(g1)-(g3) are the same as (e1)-(e3), but along cut $\#$4.
(h) The calculated band structure for single-layer FeSe with in-plane lattice constant of 3.989$\AA$.
(i) Temperature dependence of the symmetrized EDC at the Fermi crossing of $\gamma_1$ band along cut $\#$2 for FeSe$^X$. The gap is obtained following the standard fitting procedure described in ref.~\onlinecite{YZhangNP}. The original data, the fitted results and gap positions are shown in black dots, red curves, and red arrows, respectively.
(j) The superconducting gap vs. temperature for FeSe$^{X}$, compared with that of FeSe$^M$ which is reproduced from ref.~\onlinecite{FeSeTan}.}
\label{band}
\end{figure*}

After five years of intensive studies on iron-based high temperature superconductors (FeHTSs), a universal picture of the pairing symmetry has not been achieved so far. The once prevailing $s^{\pm}$ pairing \cite{Hirschfeld-review}, with the sign reversal between electron and hole Fermi surfaces, was seriously challenged by FeHTSs with only electron Fermi surfaces (called e-FeHTSs hereafter), including A$_x$Fe$_{2-y}$Se$_2$ (A= K, Cs, Rb, \textit{etc.}) \cite{ZhangNM} and single-layer FeSe on SrTiO$_3$ (STO) \cite{FeSeZhou,FeSeZhou2,FeSeTan}. For these systems, weak coupling theories based on spin-fluctuations predict a $d$-wave pairing symmetry \cite{Maier,Kreisel}. However, it is inconsistent with the isotropic superconducting gap observed by angle resolved photoemission spectroscopy (ARPES) \cite{ZhangNM,FeSeZhou,XuminPRB,Mazin1}, together with evidences for nodeless superconducting gap from specific heat \cite{HHWen}, nuclear magnetic resonance \cite{NMR}, \textit{etc.} On the other hand, the sign preserving $s$-wave pairing symmetry \cite{Hu1,Hu2,Yurong,Seo} could not account for the spin-resonance mode found in Rb$_x$Fe$_{2-y}$Se$_2$ by inelastic neutron scattering \cite{neutron}, which suggests the sign change of the superconducting order parameter on different Fermi surface sections \cite{neutronRMP}.

To explain the sign changing isotropic gap in e-FeHTSs, several novel pairing scenarios were proposed. For example, it is argued in the bonding-antibonding $s^\pm$ pairing scenario that with strong hybridization between electron pockets, the two reconstructed electron pockets can have different signs \cite{Mazin}. A further study suggested that this pairing likely coexists with the $d$-wave to form an $s+id$-wave pairing symmetry \cite{Chubukovsd}. More recently, the importance of the parity of the 2-Fe unit cell has been emphasized \cite{weik}, and it has been proposed that there are even and odd parity $s$-wave spin singlet pairing states, and the coexistence of both states gives a fully gapped state with varied signs in different Fermi surface sections \cite{Hu3,Hu4}. The hybridization between the two electron pockets is not necessary in this scenario. So far, these scenarios could not be convincingly tested, since the detailed structure of the two electron pockets could not be resolved in all known e-FeHTSs.

Two recent ARPES studies have found a gap in single-layer FeSe/STO, which closes at 65~K and suggests a possible record high superconducting transition temperature ($T_c$) of 65~K for FeHTSs \cite{FeSeZhou2,FeSeTan}; or at least, it is the pair-formation temperature record, if the superconducting transition there is a two dimensional Berezinskii-Kosterlitz-Thouless (BKT) type. Particularly, our previous ARPES study has found that the high $T_c$ in single-layer FeSe/STO is induced by suppressing the otherwise strong spin density wave (SDW) with electrons transferred from the oxygen-vacancy induced states in the substrate, and the SDW in undoped FeSe is enhanced with expanded in-plane lattice constant \cite{FeSeTan}. Consistently, the density functional theory (DFT) calculations show that it is due to the increased superexchange interactions in films under enhanced tensile strain \cite{GXG}.  We explicitly suggested that higher $T_c$  might be obtained by doping films  with further enlarged lattice  \cite{FeSeTan}, assuming the underlying spin fluctuations and superexchange interactions are important for the superconductivity.

In this paper, we have fabricated a new kind of e-FeHTS, the single-layer FeSe on top of the  Nb:SrTiO$_3$ epitaxial thin film grown on a KTaO$_3$ substrate, by successfully expanding the in-plane lattice of FeSe to 3.99$\pm$0.02$~\AA$. Our ARPES data indicate a gap closing at 70~K.  Moreover, the extremely tensile strain increases the ellipticity of the two electron Fermi surfaces at the zone corner, without any detectable hybridization. Intriguingly, the superconducting gap distribution is anisotropic with four-fold symmetry around the Brillouin zone (BZ) corner and minimal  at the crossings of the two pockets. The anisotropic but nodeless superconducting gap together with the intact electronic structure strongly supports the co-existence of the intra-pocket and inter-pocket pairing channels of the two electron pockets.  Our experiments provide important information for solving the pairing symmetry puzzle in e-FeHTSs, and also elucidate a new way to manipulate the electronic structure and enhance $T_c$ for FeSe with artificial interface.

% \textbf{The variation of anisotropy is in full agreement with two superconducting gap features observed by scanning  tunneling microscope (STM) in previous single layer materials.}

The heterostructure is designed to further enhance the tensile strain on FeSe while preserving a FeSe/Nb:SrTiO$_3$ interface (Fig.~\ref{film}(a)). KTaO$_3$ (KTO) serves as the substrate, with cubic structure and lattice constant of 3.989$\AA$ (Fig.~\ref{film}(b)), 2\% larger than that of bulk STO (3.905$\AA$). To eliminate the photoemission charging effect of KTO, silver paste was attached on the substrate edge, and 35 unit cells (u.c.) of highly conductive 5\% Nb doped STO films  \cite{NSTO} were grown layer-by-layer on KTO substrate with ozone-assisted molecular beam epitaxy (MBE) \cite{RHEEDSchlom,Supple}. Afterwards, 5~u.c. of 0.5\% Nb doped STO were epitaxially grown, with similar chemical composition as the Nb:SrTiO$_3$ substrate in the previous works  \cite{FeSeZhou,FeSeZhou2,FeSeTan}. The grown Nb:SrTiO$_3$ films were directly transferred to another MBE chamber, where FeSe thin films were grown and post-annealed following the method in ref.~\onlinecite{FeSeTan}. Details are described in the Supplementary Material \cite{Supple}. ARPES data were taken \textit{in situ} under ultra-high vacuum of $1.5\times 10^{-11} mbar$, with a SPECS UVLS discharge lamp (21.2eV He-I$\alpha$ light) and a Scienta R4000 electron analyzer. The energy resolution is 6~meV and angular resolution is 0.3$^{\circ}$. Data were taken at 25~K if not specified otherwise.

To check the actual strain on FeSe film, x-ray diffraction reciprocal lattice map was performed on the grown heterostructure around the (103) Bragg reflections. The in-plane reciprocal vector (Q$_{//}$) of the Nb:SrTiO$_3$ film equals that of KTO substrate (Fig.~\ref{film}(c)). Moreover, as shown in Fig.~\ref{film}(d), based on the high symmetry points of the photoemission intensity maps, a clear expansion of the BZ size with increasing FeSe thickness can be identified. The in-plane lattice constants of FeSe films were calculated by inversing the BZ size and plotted in Fig.~\ref{film}(e), demonstrating a rapid relaxation of the in-plane lattice in multilayer FeSe \cite{FeSeTan}. Indeed, the lattice constant of single-layer FeSe is 3.99$\pm$0.02$\AA$, which is extremely expanded. The relaxation of in-plane lattice in multi-layer films is much more rapid here than that in FeSe/Nb:SrTiO$_3$ (Fig.~\ref{film}(e)), and the cross-shape Fermi surface at the BZ corner in films with more than 2 monolayers (ML) thick is a hallmark of the SDW (Fig.~\ref{film}(d)), as shown in our previous studies \cite{FeSeTan}. Hereafter, we refer FeSe$^{X}$ to the extremely tensile strained 1ML FeSe studied in this paper, while refer FeSe$^M$  to the moderately expanded 1ML FeSe on Nb:SrTiO$_3$ substrate.

Detailed electronic structure of  FeSe$^{X}$ is studied. As shown by the photoemission intensity maps (Fig.~\ref{band}(a)), the Fermi surface of  FeSe$^{X}$ also consists of only electron pockets, similar to A$_x$Fe$_{2-y}$Se$_2$ and  FeSe$^M$. However, instead of the nearly circular and highly degenerate pockets in  FeSe$^M$, two elliptical Fermi-surfaces perpendicular to each other are observed in Fig.~\ref{band}(a), and sketched in Fig.~\ref{band}(b). Based on the Fermi surface volume, the estimated carrier concentration is 0.12 e$^-$ per Fe for  FeSe$^{X}$, similar to that of  FeSe$^M$. This indicates that strain modifies the Fermi-surface shape without much change of the charge transfer from the substrate. DFT calculations were performed on free-standing monolayer FeSe with extra 0.12 e$^-$ per Fe for the two lattice constants (Fig.~\ref{band}(c)), showing that the expanded lattice increases the ellipticity of the the electron pockets at M, which is qualitatively consistent with our experimental findings.

As shown in Fig.~\ref{band}(d1)-(d2), a parabolic band (assigned as $\alpha$) below E$_F$ can be identified around the zone center, with band top at about -72~meV. The $\omega$ band is clearly resolved around 0.21~eV below E$_F$ near $\Gamma$, which is often observed in iron pnictides with $d_{z^2}$ orbital character. Around the zone corner (Fig.~\ref{band}(e1),(f1)), the $\gamma$ band in  FeSe$^M$ \cite{FeSeZhou} splits into $\gamma_1$ and $\gamma_2$ bands in  FeSe$^{X}$, due to the lifted degeneracy of $\gamma_1$ and $\gamma_2$ pockets. In this 2-Fe BZ, the $\gamma_1$ ($\gamma_2$) band is intense around M1 (M2), while its folded band around the neighboring BZ corner M2 (M1) is weak in intensity but can still be tracked in MDCs (momentum distribution curves) as shown in Figs.~\ref{band}(f2) (Figs.~\ref{band}(e2)). By fitting the MDC at E$_F$ with four Lorentzian peaks (Fig.~\ref{band}(e3),(f3)), the Fermi wave-vector ($k_F$) is resolved, which is 0.51~$\AA^{-1}$ for the major axis and 0.37~$\AA^{-1}$ for the minor axis of the elliptical pocket. The deduced $k_F$'s from cut $\#$2 and cut $\#$3 are consistent. Figure~\ref{band}(g1) shows the photoemission intensity along cut $\#$4 where $\gamma_1$ and $\gamma_2$ pockets intersect. Remarkably, the MDCs (Fig~\ref{band}(g2)-(g3)) show a single-Lorentzian-peak behavior for both sides, without any hybridization-induced band anti-crossing. Normally, this is only expected for a free-standing single-layer FeSe,  as illustrated by our DFT calculations in Fig~\ref{band}(h), when there is  no interlayer hopping and the $S_4$ symmetry is preserved. The negligible hybridization here thus suggests that the low energy electronic structure of the FeSe layer remains intact without much influence from the substrate. The screening from the substrate phonon on the Cooper pairing in FeSe  thus should be unlikely  \cite{DHLee}. Upon post-annealing under vacuum and Se flux \cite{Supple}, the superconducting gap size and sample quality are tuned. As shown in Fig.~\ref{band}(i), one could still observe the signature of gap in the 65~K data and in the corresponding fit \cite{YZhangNP}. The temperature dependence of the gap can be well-fitted by the BCS gap temperature dependence function in Fig.~\ref{band}(j). These suggest that this film has a possible $T_c$ of 70~K, assuming the gap is not due to Cooper pair pre-formation in the normal state. The $T_c$ is slightly enhanced in this film compared with  FeSe$^M$(Fig.~\ref{band}(j)), probably due to the enhanced superexchange interactions with increased lattice constant here \cite{FeSeTan,GXG}.

\begin{figure}[t]
\includegraphics[width=8.6cm]{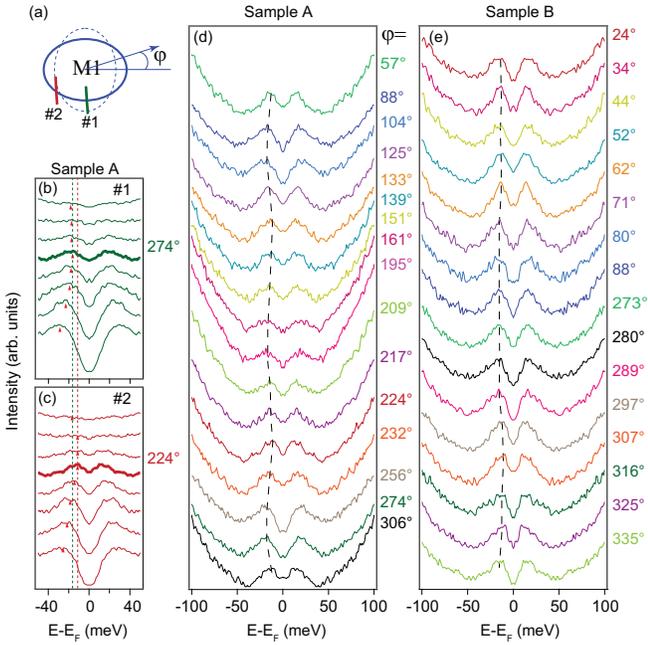}
\caption{(color online). (a) Fermi surface sheets around M1 and the definition of cut $\#$1, cut $\#$2, and polar angle $\varphi$.
(b),(c) Symmetrized EDCs for sample A along cut $\#$1 and $\#$2, respectively. The $k_F$'s of $\gamma_1$ are shown by the thicker curves, with the polar angle of 274$^\circ$ and 224$^\circ$, respectively.
(d) Symmetrized EDCs at $k_F$'s of $\gamma_1$ band for sample A, with the momentums counterclockwise along $\gamma_1$ pocket as indicated by the polar angles.
(e) is the same as panel (d), but measured on sample B.
} \label{gapani}
\end{figure}

\begin{figure}[t]
\includegraphics[width=8.6cm]{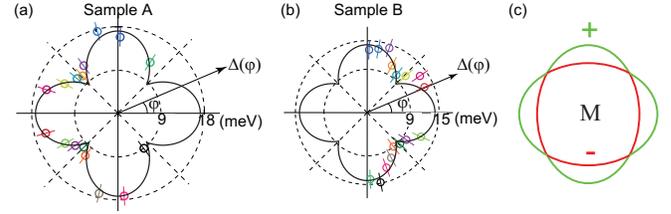}
\caption{(color online). (a) Gap distribution of the $\gamma_1$ pocket in polar coordinates for sample A, where the radius represents the gap size, and the polar angle represents $\varphi$. The gap is estimated through an empirical fit  \cite{YZhangNP}, and the error bars come from the standard deviation of the fitting process. Solid black curve shows the fitting by $\Delta$=$\Delta_0+\Delta_1 |\cos(2\varphi)|$, with $\Delta_0$=8.52~meV and $\Delta_1$=8.69~meV.
(b) same as panel (a), but measured on Sample B. The black curve shows the fitting results with $\Delta_0$=8.41~meV and $\Delta_1$=5.73~meV. (c) Cartoon for the sign-changing s-wave pairing symmetry around M in a recent model, and it is arranged in  $A_{1g}$ symmetry on the four zone corners around $\Gamma$ \cite{Hu3,Hu4}.
} \label{sum}
\end{figure}

The momentum dependence of the superconducting gap is further investigated. Figures~\ref{gapani}(b) and \ref{gapani}(c) show the EDCs (energy distribution curves) along the two cuts indicated in Fig.~\ref{gapani}(a). The EDCs are symmetrized with respect to E$_F$ to remove the influence of Fermi-Dirac cut-off. Due to the low intensity of the folded Fermi surface, the superconducting gap of $\gamma_1$ band can be deduced without interference from the $\gamma_2$ band. At the $k_F$, the spectra lose half intensity and the EDCs bend back. The vertical green and brown dashed lines indicate the coherence peak positions at two $k_F$'s  (274$^{\circ}$ and 224$^{\circ}$ in the polar coordinates) respectively, and they unambiguously differ from each other. The symmetrized spectra in the superconducting state at various $k_F$'s of $\gamma_1$ pocket are shown in Fig.~\ref{gapani}(d). The peak positions of the coherence peaks differ at different momenta, indicating an anisotropy in superconducting gaps. To quantify the anisotropy, the superconducting gap size are fitted \cite{YZhangNP}, and plotted in a polar coordinate in Fig.~\ref{sum}(a). Beyond the finite error bars, a clear anisotropy of superconducting gap can be recognized, following four-fold symmetry. The superconducting gap distribution on another sample is shown in Fig.~\ref{gapani}(e) and Fig.~\ref{sum}(b). Despite of the varied superconducting gap size due to different post-annealing process, anisotropic but nodeless superconducting gap distribution along the electron Fermi surface is alike in different samples.

Although the in-plane anisotropic superconducting gap on the electron Fermi surfaces has not been observed in e-FeHTSs before, it has been reported in several iron pnictides.  For example, we recently  found that the superconducting gap is anisotropic in NaFe$_{0.9825}$Co$_{0.0175}$As due to the coexistence of SDW and superconductivity, with its minima at $\varphi$=0 and $\pi/2$  \cite{QQGPRX}. In addition, gap minima along $\varphi$=45$^{\circ}$ were reported for an electron pocket of LiFeAs, which is attributed to band hybridization and the mixture of  $(\cos k_x+\cos k_y )$-term in the gap function  \cite{LiFeAs}. However, none of these could account for the over 50$\%$ variation of the gap size in the elliptical Fermi surface of FeSe$^{X}$ as neither SDW nor  hybridization is present.

The observed nodeless superconducting gap poses strong constraints on theoretical scenarios. Based on the $d$-wave pairing, if the electron pockets are elliptical with negligible hybridization, gap nodes would be induced in the folder BZ, which is not observed here. The bonding-antibonding $s^\pm$  \cite{Mazin} or  $s+id$ symmetry  \cite{Chubukovsd} also can not explain our observations, since the required sizable hybridization between the electron pockets is absent in the data. We note that our data might be consistent with the theory with even and odd parity $s$-wave spin singlet pairing channels \cite{Hu3}. This theory suggests that a full gap with a sign change on the electron pockets can be realized without the hybridization by combining the two different $s$-wave pairings: the even parity $s$-wave pairing contributes a $d$-wave-like momentum dependence of the gap structure around each electron pocket, $\Delta_e \sim cos(2\varphi)$, with nodes around $\varphi=\pm\pi/4, \pm3\pi/4$ as sketched in Fig.~\ref{sum}(c);  and the odd parity one causes the pair coupling between the two even parity ones on the two electron pockets, and lifts the nodes with a gap determined by the odd parity pairing  $\Delta_o$.  The convolution of these two momentum-dependent pairing components generally produces an anisotropic gap function \cite{Hu4}.  In Figs.~\ref{sum}(a) and \ref{sum}(b), we show that the experimental gap can be  well fitted by $\Delta$=$\Delta_0+\Delta_1 |\cos(2\varphi)|$, which agrees with the above theory if taking $\Delta_o=\Delta_0$ at  $\varphi=\pm\pi/4, \pm3\pi/4$, and $\Delta_e\approx\Delta_0+\Delta_1$ at $\varphi=0, \pm\pi/2$, and $\pi$ according to the theory \cite{Hu3,Hu4}.

To summarize, we have shown that   heterostructure  provides a novel path  for  exploring  superconductivity in FeHTSs, and revealed the novel electronic structure and gap distribution in extremely tensile strained single-layer FeSe.  The lifted degeneracy of the electron pockets with negligible hybridization, together with the anisotropic but nodeless superconducting gap, provide important experimental foundations for solving the pairing symmetry puzzle of e-FeHTSs.

\textit{Acknowledgements:} We gratefully acknowledge the fruitful discussions with Prof. Jianxin Li and Prof. Jian Shen, and we thank Prof. Dawei Shen and Prof. Wei Peng for helping with x-ray diffraction measurements. This work is supported in part by the National Science Foundation of China, and National Basic Research Program of China (973 Program) under the grant No. 2012CB921402.

\end{document}